\documentclass{article}

\usepackage{titling}
	\newcommand*{\thetitle}{Cloud Brokerage: A Systematic Survey}
\usepackage{amsfonts,amsmath,amssymb,amstext,latexsym,pifont}
\usepackage{adjustbox}
\usepackage{array,makecell}
\usepackage[inline]{enumitem}
	\setlist[enumerate]{leftmargin=3em,itemsep=0.33em}%
	\setlist[itemize]{leftmargin=3em,label={$\bullet$}}%
\usepackage{graphicx}
\usepackage{hyperref}
\usepackage{natbib}
	\setcitestyle{square,comma,numbers,sort&compress}
\usepackage[sc]{mathpazo} 
\usepackage[T1]{fontenc} 
\usepackage{microtype} 
\usepackage[hmarginratio=1:1,top=25mm,columnsep=20pt]{geometry} 
\usepackage[hang, small,labelfont=bf,up,textfont=it,up]{caption} 
\usepackage{booktabs} 
\usepackage{float} 
\usepackage{abstract} 

\usepackage{titlesec} 
\renewcommand\thesubsection{\Roman{subsection}} 
\titleformat{\section}[block]{\large\scshape\centering}{\thesection.}{1em}{} 
\titleformat{\subsection}[block]{\large}{\thesubsection.}{1em}{} 

\usepackage{fancyhdr} 
\usepackage{color}
\usepackage{colortbl}
\usepackage[usenames,dvipsnames,table]{xcolor}
\usepackage{tikz}
\usepackage{xspace}
\usepackage{booktabs}

\usepackage{pifont}
\newcommand{\cmark}{{\color{NavyBlue}\ding{51}}}
\newcommand{\xmark}{{\color{Black!40}\ding{53}}}

\newcommand*{\cf}{\textit{cf.}\@\xspace}
\newcommand*{\eg}{\textit{e.g.,}\@\xspace}
\newcommand*{\ie}{\textit{i.e.}\@\xspace}
\makeatletter
\newcommand*{\etc}{%
    \@ifnextchar{.}%
        {etc}%
        {etc.\@\xspace}%
}
\newcommand*{\etal}{%
    \@ifnextchar{.}%
        {et al}%
        {et al.\@\xspace}%
}
\makeatother

\newcolumntype{R}[2]{%
    >{\adjustbox{angle=#1,lap=\width-(#2),raise=(#2)}\bgroup}%
    l%
    <{\egroup}%
}
\newcommand*\rot{\multicolumn{1}{R{45}{1em}}}%
\newcommand*\vrt{\multicolumn{1}{R{90}{0em}}}%

\definecolor{Gray}{gray}{0.9}
\newcolumntype{g}{>{\columncolor{Gray}}c}

\usepackage{etoolbox}

\makeatletter
\patchcmd{\@classz}
  {\CT@row@color}
  {\oldCT@column@color}
  {}
  {}
\patchcmd{\@classz}
  {\CT@column@color}
  {\CT@row@color}
  {}
  {}
\patchcmd{\@classz}
  {\oldCT@column@color}
  {\CT@column@color}
  {}
  {}
\makeatother

\usepackage[ruled]{algorithm2e}

\SetAlFnt{\small}
\SetAlCapFnt{\small}
\SetAlCapNameFnt{\small}
\SetAlCapHSkip{0pt}
\IncMargin{-\parindent}

\begin{document}

\date{}
\title{\thetitle}

\author{Abdessalam Elhabbash, Faiza Samreen, James Hadley, Yehia Elkhatib\\[4mm]
\normalsize Email: \{i.lastname@lancaster.ac.uk\}\\[4mm]
\small MetaLab, School of Computing and Communications, Lancaster University, UK\\
}

\maketitle

\thispagestyle{fancy}

\begin{abstract}
\textit{Background}--- The proliferation of cloud providers and provisioning levels has opened a space for cloud brokerage services. Brokers intermediate between cloud customers and providers to assist the customer in selecting the most suitable cloud service, helping to manage the dimensionality, heterogeneity, and uncertainty associated with cloud services. 
\textit{Objective}--- This paper identifies and classifies approaches to realise cloud brokerage. By doing so, this paper presents an understanding of the state of the art and a novel taxonomy to characterise cloud brokers.
\textit{Method}--- We conducted a systematic literature survey to compile studies related to cloud brokerage and explore how cloud brokers are engineered. We analysed the studies from multiple perspectives, such as motivation, functionality, engineering approach, and evaluation methodology. 
\textit{Results}--- The survey resulted in a knowledge base of current proposals for realising cloud brokers. 
The survey identified surprising differences between the studies' implementations, with engineering efforts directed at combinations of market-based solutions, middlewares, toolkits, algorithms, semantic frameworks, and conceptual frameworks.
\textit{Conclusion}--- Our comprehensive meta-analysis shows that cloud brokerage is still a formative field. There is no doubt that progress has been achieved in the field but considerable challenges remain to be addressed. This survey identifies such challenges and directions for future research.
\end{abstract}

\textbf{Keywords:} Cloud computing; Cloud brokerage; Systematic literature review; Survey.

\section{Introduction}
\label{sec:intro}
Despite the promised benefits of cloud computing (such as low cost, high availability, and flexible application deployment), the adoption of cloud brokers in practice remains sparse \cite{Satzger2013,Elkhatib2016crosscloudmap,Eisa2016}. The market for cloud services is overwhelmed with a high number of heterogeneous cloud offerings, making the selection of a cloud service a challenging task for the cloud services customer (CSC) \cite{etsi-csc-wp1,appdirect2017,Elkhatib2018same,ghrada2018}. Furthermore, considering that each cloud service provider (CSP) exposes their unique API, designing and developing an application so that it can be deployed on a specific CSP does little to mitigate the development efforts to move the application, \eg if it performs badly, on the selected CSP. 

To bridge the above gaps, the cloud community has long proposed that cloud brokers intermediate between CSCs and CSPs, thereby mitigating the risk of selecting any given CSP. Brokerage can benefit the CSCs by abstracting away differences between CSPs and by helping CSCs to find the most suitable cloud services. These benefits aim to enable CSCs to lower costs and seamlessly switch between CSPs to ensure that their application requirements are always met. These benefits can also encourage forgoer businesses to trust and thus adopt cloud computing \cite{ANASTASI20171}. Additionally, cloud brokers can benefit CSPs by helping them to exploit economies of scale and by offloading some of their support and management overheads to the brokers \cite{Ferrer201266}. Cloud brokers can also benefit CSPs by reducing energy consumption through aggregation and consolidation \cite{srikantaiah2008energy,QUARATI2013}.

To realise brokerage, applications must be able to cross the boundaries of any given CSP. This gives rise to the field of \emph{cross-cloud computing} \cite{Elkhatib2016crosscloudmap} that aims to support application developers with the challenges associated with interoperability, resource scheduling strategies, dynamic deployment, and migration. Within this growing field, the term `brokerage' has been used to refer to different intermediation models. One such model is that of cloud federation, mandating that competing CSPs agree and implement common technologies (such as virtualisation technologies and APIs) to enable the broker to select a CSP service, deploy use CSC's application and adapt deployments. In contrast, the multi-cloud model does not assume common technologies. Instead, to facilitate switching between CSPs, the broker's role must also encompass abstracting away the differences between CSPs. A third model is the decision support system, in which brokerage is limited to recommending a cloud service to CSCs based on their requirements. In this model, the practical aspects of deploying and relinquishing services are left to the CSC.

The breadth of challenges surrounding cloud brokerage has brought about so much research there that is a need to systematically analyse the proposed solutions. A systematic literature survey (SLR) is a methodological survey that aims to systematically capture efforts relating to a specific topic to develop a comprehensive and unbiased knowledge base around the topic~\cite{kitchenham2004procedures}. This paper provides the results of an SLR that was carried out on the topic of cloud brokerage. The SLR results are analysed and presented as part of a taxonomy that will aid researchers in this field to appreciate both the big picture and some of the finer details of work in this area. In this sense, this paper aims to review and analyse existing solutions in terms of:
\begin{itemize}
    \item exploring the reasons that motivate the need for cloud brokers; 
    \item developing a fundamental understanding of the tasks carried out by cloud brokers; 
    \item investigating approaches to engineer and realise cloud brokers; 
    \item examining approaches to evaluate proposed solutions; and 
    \item identifying the main limitations of current solutions and highlighting areas for future research where improvements can be made. 
\end{itemize} 

The remainder of this paper is organised as follows. The next section briefly presents other reviews related to cloud brokerage. Section \ref{sec:meth} describes the methodology followed to conduct this survey including the objectives and research questions. Section \ref{sec:reporting} gives an overview of the selected papers. Sections \ref{sec:moivation}, \ref{sec:functionalities}, and \ref{sec:rq3} answers the research questions. Section \ref{sec:chr} presents a characterisation framework of the selected papers to synthesise and discuss the results and identify the limitations of current solutions. Section \ref{sec:future} identifies future areas of research. Section \ref{sec:discussion_limitations} reflects on the results and discusses threats of validity of this SLR. Finally section \ref{sec:conc} concludes the article. 

\section{Related Work}
\label{sec:rw}

To the best of our knowledge, there is no other comprehensive or systematic survey of works on cloud brokerage. However, we briefly present some reviews that relate at varying degrees to cloud brokerage.  

\citet{Eisa2016} surveys some cloud services selection approaches. The authors focus on analysing three commercial CSP search tools that help CSCs to search for cloud services. The survey also presents a number of academic works related to cloud selection. In their conclusions, the authors report the need for cloud brokers that can extend and use cloud services selection tools to assist CSCs. Other similar surveys also focus on cloud service selection, \eg \cite{SUN2014134}.

\citet{DBLP:journals/corr/BarkerVT15} reviewed prominent commercial solutions in cloud brokerage from an academic perspective, classified them into one of four categories (performance, migration, theoretical models and data), and outlined a research roadmap in light of these efforts. \citet{Grozev2014} analysed a number of the early cross-cloud application brokering mechanisms, while \citet{ASSIS201651} focused on cloud federations.

There is a limited number of surveys on interoperability and portability issues in the cloud. These studies tend to focus on the interoperability challenge from a particular angle: 
\citet{Loutas2011Interoperability} is focused predominantly on semantic divergence in the cloud ecosystem as a root cause for the interoperability challenge.
\citet{Zhang2013interop} presents a high-level taxonomy of issues relating to interoperability at the IaaS level, encompassing broad issues ranging from APIs and GUIs to virtualisation technologies, encryption mechanisms and SLA verification. \citet{Kaur2017} surveys and analyses approaches implementing interoperability and portability in different inter-cloud models. Although their work shares some similarities with ours, our analysis is more from the CSC's perspective (\ie how the brokers benefit the CSCs) whereas their analysis is more from the CSP's perspective (\ie the interoperability and collaboration between the CSPs). 

There have been surveys on other aspects of cloud computing, such as design issues (\cf~\cite{Zhang2010soa}), resource management (\cf~\cite{Manvi2014mng,Zhan2015scheduling}), monitoring (\cf~\cite{Aceto2013monitoring}), migration (\cf~\cite{Jamshidi2013migration}), service composition (\cf~\cite{Jula2014composition}), security (\cf~\cite{Iankoulova2012security,Patel2013ids}), elasticity~\cite{al2017elasticity}, 
among other subjects.

\section{Survey Methodology}
\label{sec:meth}

\begin{figure}[hbt]
\centering
\includegraphics[width=\textwidth]{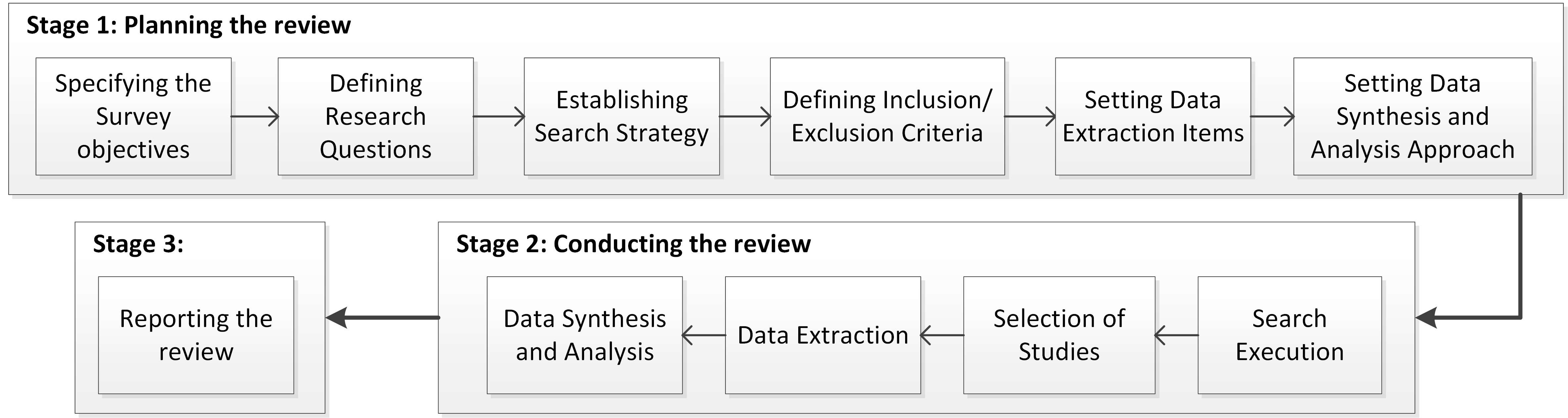}
\caption{Research methodology.}
\label{fig:meth:methodology}
\end{figure}

The purpose of this study is to comprehensively survey the literature to identify the state of the art when it comes to brokerage in cloud computing. 
This field is a vast and growing one: thousands of papers have been published on many different aspects of cloud computing and by researchers from different backgrounds (\eg distributed computing, high performance computing, grid computing, optimisation theory, financial derivatives, \etc). As such, there is considerable variance in the terminology used in the literature which is supported both by previous work (\cf \cite{Elkhatib2016crosscloudmap}) and by our preliminary survey. 
Adopting a systematic surveying approach is thus the most rigorous way of identifying all relevant work produced by researchers across disciplines. \figurename~\ref{fig:meth:methodology} depicts the methodology followed to conduct this study.

\subsection{Research Objectives}
Our objectives are as follows.
\begin{enumerate}[label=\texttt{O{\arabic*}}]
  \item Build a library of work that is related to the general topic of cloud brokerage, and provide this to other researchers as an open dataset.
  \item Identify a more focused set of works that have shaped research in cloud brokerage.
  \item Characterise existing solutions in the field of cloud brokerage, and clarify the similarities and differences between them using a characterisation framework.
  \item Produce a taxonomy of the state of the art to further highlight the adopted approaches and methods.
  \item Distinguish gaps in the state of the art in terms of research challenges and approaches.
\end{enumerate}

\subsection{Research Questions}
We conduct this study by addressing the following research questions.
\begin{enumerate}[label=\texttt{RQ{\arabic*}}]
  \item \emph{What is the motivation for designing cloud brokers?}\\
  We address the need to identify the requirements that have stimulated the research and development of cloud brokerage systems. 
  
  \item \emph{What are the functionalities of a cloud broker?}\\
  We identify the range of operations that are performed by cloud brokers in order to achieve their goals. 
  
  \item \emph{What are the approaches for engineering cloud brokers?}\\
  We identify engineering approaches used to implement cloud brokers and investigate links between approaches and outcomes.
\end{enumerate}

\subsection{Search and Selection Strategy}
Cloud computing is obviously an extremely active field. It is for this reason that our survey method had to empirically draw related work from all relevant sources.

For work from the research community, we built a knowledge base from seven major publishers and online databases in computer science: ACM DL, IEEExplore, ScienceDirect, SpringerLink and Wiley Online Library. 
We captured works from these sources using a rigorous \textbf{five-phase procedure} that is described below. 
Aggregate indexing services such as Scopus and Web of Science were also consulted but eventually disregarded as they had only duplicated works harvested from the original publisher databases. 

\subsubsection{Search Phase}

In the first phase, we query the main publication portals using a set of different search terms. The retrieved results along with their metadata (including title, abstract, and publication outlet) are stored in a local knowledge base which will be filtered in the subsequent phases. 

\begin{figure}
\centering
\def\layersep{2cm}
\begin{tikzpicture}[pin distance=0mm, draw=gray!50, node distance=\layersep]
    \tikzset{every pin edge/.style={draw=white, ultra thin}}
    \tikzstyle{neuron}=[circle,minimum size=15pt,inner sep=1pt]
    \tikzstyle{primary}=[neuron,fill=green!30];
    \tikzstyle{secondary}=[neuron,fill=red!40];
    \tikzstyle{annot} = [text width=2cm, text centered]
    
    \node[primary, pin=left:broker] (p1) at (0,-3) {};
    \node[primary, pin=left:brokerage] (p2) at (0,-5) {};

    \node[secondary, pin=above:cloud] (s1) at (3.0\layersep,-1 cm) {};
    \node[secondary, pin=above:cross-cloud] (s2) at (3.0\layersep,-2 cm) {};
    \node[secondary, pin=above:federated cloud] (s3) at (3.0\layersep,-3 cm) {};
    \node[secondary, pin=above:inter-cloud] (s4) at (3.0\layersep,-4 cm) {};
    \node[secondary, pin=above:hybrid cloud] (s5) at (3.0\layersep,-5 cm) {};
    \node[secondary, pin=above:multi-cloud] (s6) at (3.0\layersep,-6 cm) {};
    \node[secondary, pin=above:iaas] (s7) at (3.0\layersep,-7 cm) {};

    \node[secondary, pin=right:unified] (t1) at (6.0\layersep,-2 cm) {};
    \node[secondary, pin=right:holistic] (t2) at (6.0\layersep,-3 cm) {};
    \node[secondary, pin=right:optimised] (t3) at (6.0\layersep,-4 cm) {};
    \node[secondary, pin=right:comprehensive] (t4) at (6.0\layersep,-5 cm) {};
    \node[secondary, pin=right:heterogeneous] (t5) at (6.0\layersep,-6 cm) {};

    \foreach \source in {1,...,2}
        \foreach \dest in {1,...,7}
            \path (p\source) edge (s\dest);
    \foreach \source in {1,...,7}
        \foreach \dest in {1,...,5}
            \path (s\source) edge (t\dest);

    \node[annot,above of=s1, node distance=1.5cm] (sec-label) {\bf Secondary keywords};
    \node[annot,left of=sec-label, node distance=3.5cm] {\bf Primary keywords};
    \node[annot,right of=sec-label, node distance=3.5cm] {\bf Additional keywords};

\end{tikzpicture}
\caption{Search queries used to identify works to include in our knowledge base}
\label{tab:meth:searchcombos}
\end{figure}
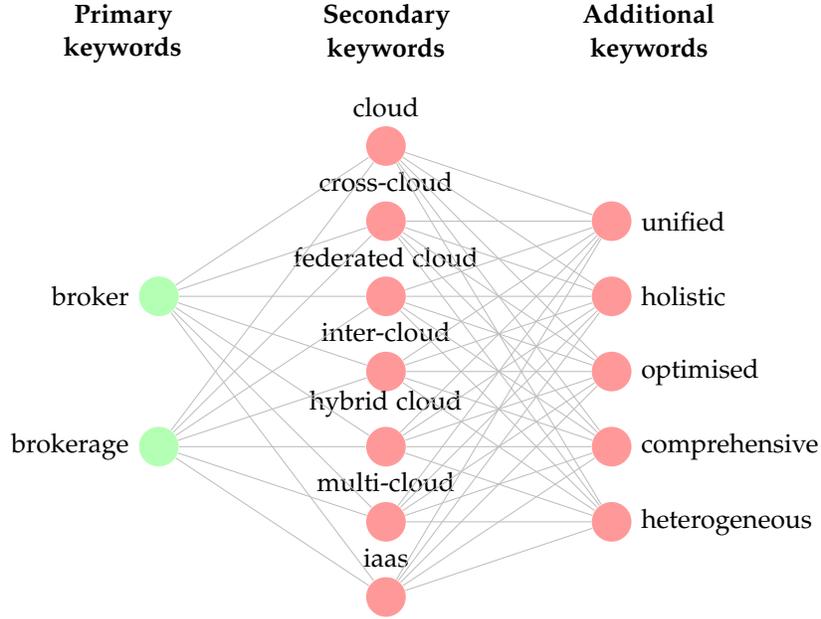

The search string combinations used are outlined in \figurename~\ref{tab:meth:searchcombos}. The primary keywords were chosen as key identifiers of work in the area of cloud brokerage. 
However, cloud brokerage is a relatively young and developing field~\cite{Elkhatib2016crosscloudmap}, 
and as such we needed to augment these keywords with others that are analogous. 
The second level keywords were therefore added to capture any publications that tackle the heterogeneity challenge albeit not under the label of brokerage. 
Alternative spellings (\eg ``inter-cloud'', ``intercloud'', and ``inter cloud'', American and British spelling variations, \etc) were included to ensure comprehensive coverage.
Primary keywords ($K_P$) were used with any of the secondary ($K_S$) or additional ($K_A$) keywords; \ie $\forall K_P \: \land \: \forall K_S \: \land \: \forall K_A$.

Finally, we augmented this with keywords extracted from well known works on cloud brokers that we were familiar with before carrying out the survey (such as~\cite{Ferrer201266,Tordsson2012brokering}). These are shown on the far right in the diagram.

\subsubsection{Screening Phases and Selection Criteria}
\label{subsubsec:screening}
The subsequent three phases filter the knowledge base based on the contents of the publication title, abstract, introduction and conclusion sections, respectively. 
Authors and publishing outlet (journal, conference, \etc) were intentionally left out to avoid bias towards publications based on their provenance and to ensure that relevance was judged purely on the content conveyed by the paper's metadata and in key sections. We considered that not all titles and abstracts are truly reflective of the contents of a paper. However, we concluded that our approach was necessary to avoid provenance-based preconceptions. 

A large number of works reduce brokerage to a mere resource allocation problem, \eg~\cite{Mehrotra2016195}, 
and as such assume cooperation between CSPs to gloss over interoperability and comparability challenges. Other studies (\eg \cite{Javed201652}) have a strong expectation of active involvement of cloud providers in issuing tenders or responding to calls for bids. We also disregard these as history has shown us that CSPs are not interested in this form of marketplace \cite{Johnston2014swap}. 

Table \ref{tab:meth:selectionCriteria} summarises the inclusion and exclusion criteria that were adopted to perform the selection. Every effort was made to retrieve manuscripts behind paywalls through our institutional subscription. Of the papers that reached the third phase, only 3 were not accessible (0.03\% of the total, 1.68\% at phase 3). 

\begin{table*}[ht]
\caption{Studies Selection Criteria\label{tab:meth:selectionCriteria}}{
	\renewcommand{\arraystretch}{1.4}
\begin{tabular}{ l p{12cm} } 
\hline
&	\textbf{Inclusion Criteria}	 
\\
\hline
\textbf{I1.} &	Research papers presenting new and emerging ideas.	
\\
\textbf{I2.} &	Literature published as books, book chapters, collections, and technical reports.
\\
\textbf{I3.} &	Papers developing or extending brokerage systems in the cloud paradigm.
\\
\textbf{I4.} &	Papers discussing aspects of brokerage in the cloud paradigm. 
\\
\hline
&	\textbf{Exclusion Criteria}
\\
\hline
\textbf{E1.} &	Papers not in the form of a full research paper, i.e. in the form of abstract, tutorials, presentation, or essay.
\\
\textbf{E2.} &	Papers without an available abstract, or behind a paywall.
\\
\textbf{E3.} &	Papers not written in the English language.
\\
\textbf{E4.} &	Papers reducing brokerage to a mere resource allocation problem or restricting to a single application type such as MapReduce. 
\\
\textbf{E5.} &	Papers focusing on brokerage in other paradigms; i.e. not the cloud paradigm.
\\
\hline
\end{tabular}
}
\end{table*}

When the same work was published in more than one outlet, as is common with conference papers that then get expanded into a journal article, only the original work is included unless there is an enhancement in the follow up publication that is significant in brokerage-related aspects. 

During each screening phase, every publication is screened by at least two of this paper's authors. 
Each screener is asked to judge the relevance of the paper and, accordingly, choose to either \textit{Accept}, \textit{Reject}, or \textit{Defer}. 
If the two screeners do not agree in their choices or if one of them chooses to defer his/her decision, then the publication in question is marked for discussion by the screeners until a decision is reached. 
In the final screening phase, the screeners also extracted keywords from the remaining set. These are used to identify the most suitable reviewer to further scrutinise the paper. 
Only papers with an original technical contribution are selected; editorials are ignored at this phase, but the most pertinent of these will be revisited in the Related Work section.

At the cost of additional deferred inspection effort, we decide to err on the side of caution and accept papers that are deemed to still be borderline even after deliberation between the screeners.

\subsubsection{Inspection Phase}

The final phase is to thoroughly read the accepted publications to ascertain their relevance and contribution in the field of cloud brokerage. The papers are categorised according to their contributions to achieve a systematic knowledge base which is the core goal of this study.

\section{Reporting the review}
\label{sec:reporting}

This section provides an overview of the selected studies. 

\subsection{Overview of the intermediate selection process outcome}

We present the high-level results of the SLR. 
Table~\ref{tab:meth:phases} presents a concise quantitative summary of the results of our search, screening and inspection phases across the 7 publication libraries. 
The survey captured a very large number of papers. Screening phases helped reduce this number but at a much lower rate than we expected. This is mainly because many papers promise a brokerage solution when in fact they offer a supporting technology or model. 
A representative example of this is CloudCmp~\cite{Li2010cloudcmp}, an early system to compare multiple cloud providers in terms of performance and cost. However, in essence, CloudCmp is a standard benchmarking suite for Java virtual machines. Other work followed using similar approaches, \eg \cite{6008753,AliSPE1072}. 
Such work could certainly be used to support the construction of a broker but does not in itself implement the full range of services typically associated with a broker. 
Other examples include scheduling optimisations, negotiating algorithms, and ontologies. 

\begin{table*}[th]
\setlength{\tabcolsep}{4pt}
\caption{Selection phases and results.\label{tab:meth:phases}}{
	\renewcommand{\arraystretch}{1.4}
	\begin{tabular}{rl>{\raggedright}p{3.7cm}rrrrrrrr}
 		Phase & Process & Selection criteria & \rot{ACM DL} & \rot{arXiv.org} & \rot{IEEExplore} & \rot{PLOS} & \rot{ScienceDirect} & \rot{SpringerLink} & \rot{Wiley Online Library} & \rot{\textbf{Total}} \\
    \toprule
1& Search & Keywords (\figurename~\ref{tab:meth:searchcombos}) & 3517 & 606 & 502 & 21 & 1574 & 4413 & 214 & \textbf{10847} \\
2& Screening & Title & 101 & 74 & 156 & 2 & 142 & 190 & 16 & \textbf{681} \\
3& Screening & Abstract & 34 & 18 & 42 & 0 & 44 & 36 & 5 & \textbf{179} \\
4& Screening & Introduction and Conclusion & 17 & 11 & 20 & 0 & 27 & 19 & 4 & \textbf{98} \\
5& Inspection & Full paper & 9 & 2 & 4 & 0 & 10 & 6 & 2 & \textbf{33} \\
    \bottomrule
	\end{tabular}
}
\end{table*}

\subsection{Overview of the selected studies}

\begin{figure}[hbt]
\centering
\includegraphics[width=0.51\textwidth]{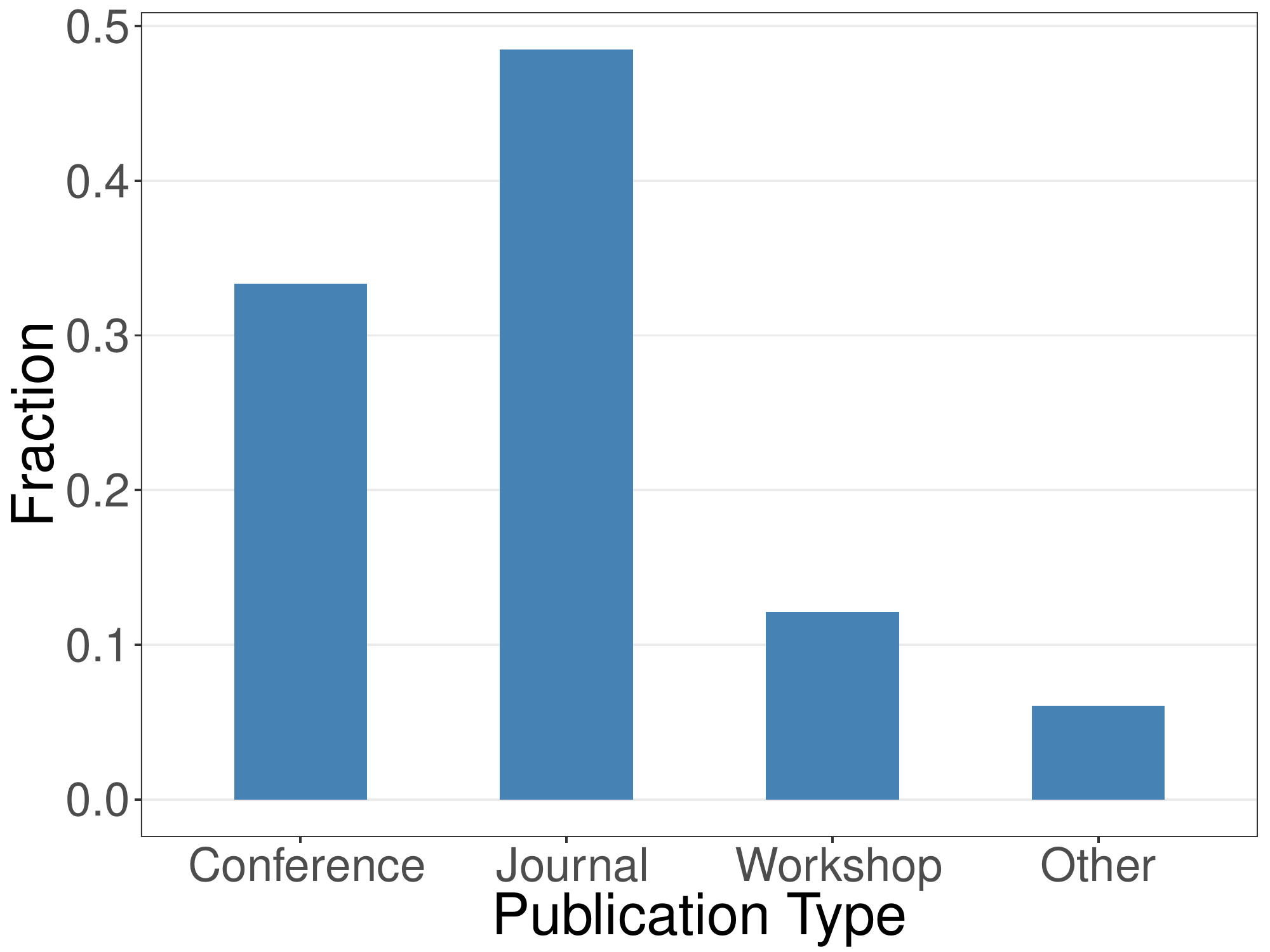}
\caption{Total distribution of publication types of selected studies.}
\label{fig:results:pubtypes}
\end{figure}

\begin{figure}[htb]
\centering
\includegraphics[width=0.8\textwidth]{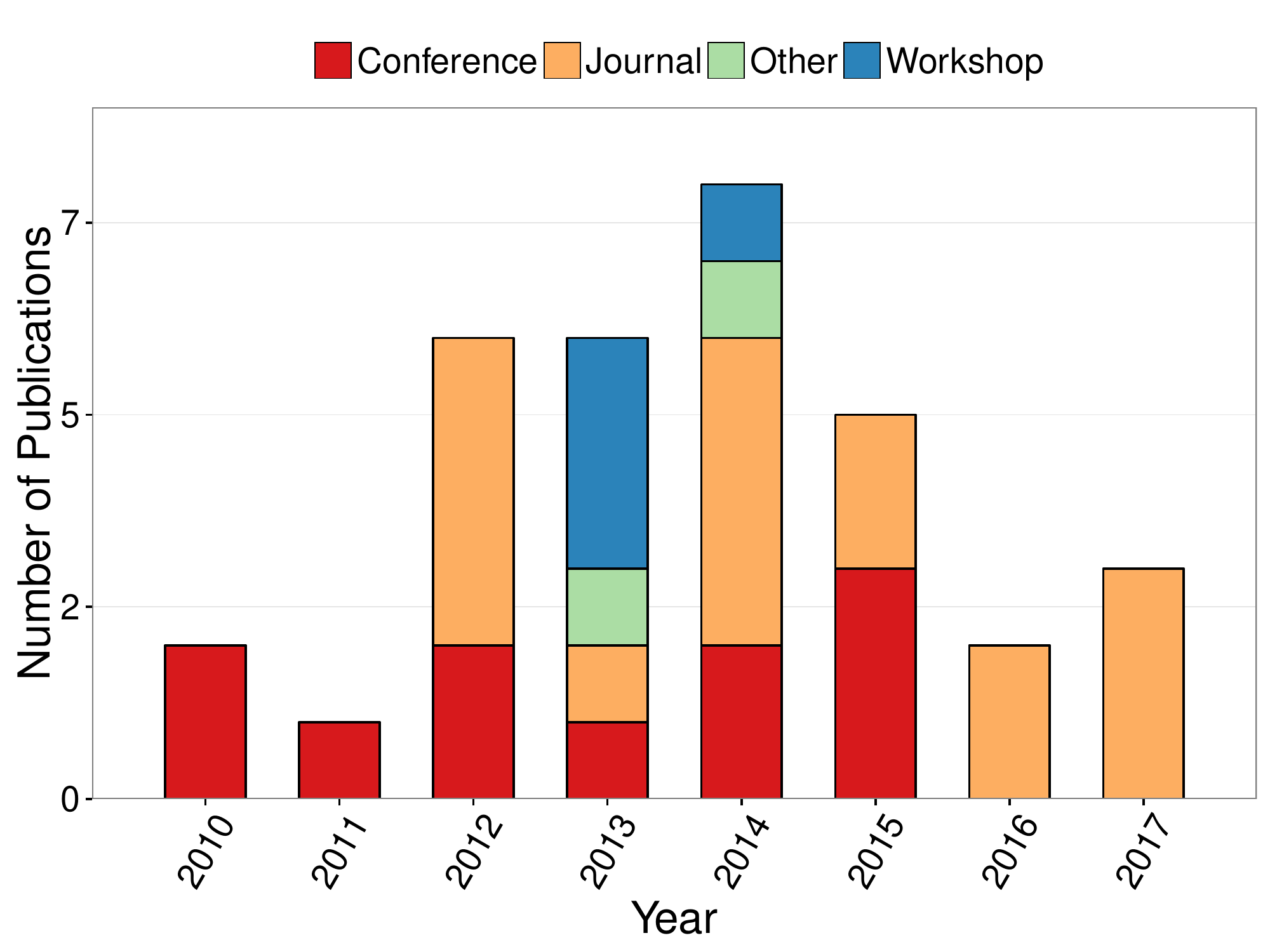}
\caption{Distribution of selected studies by type over publication year.}
\label{fig:results:pubyear}
\end{figure}

\begin{figure}[htb]
\centering
\includegraphics[width=0.8\textwidth]{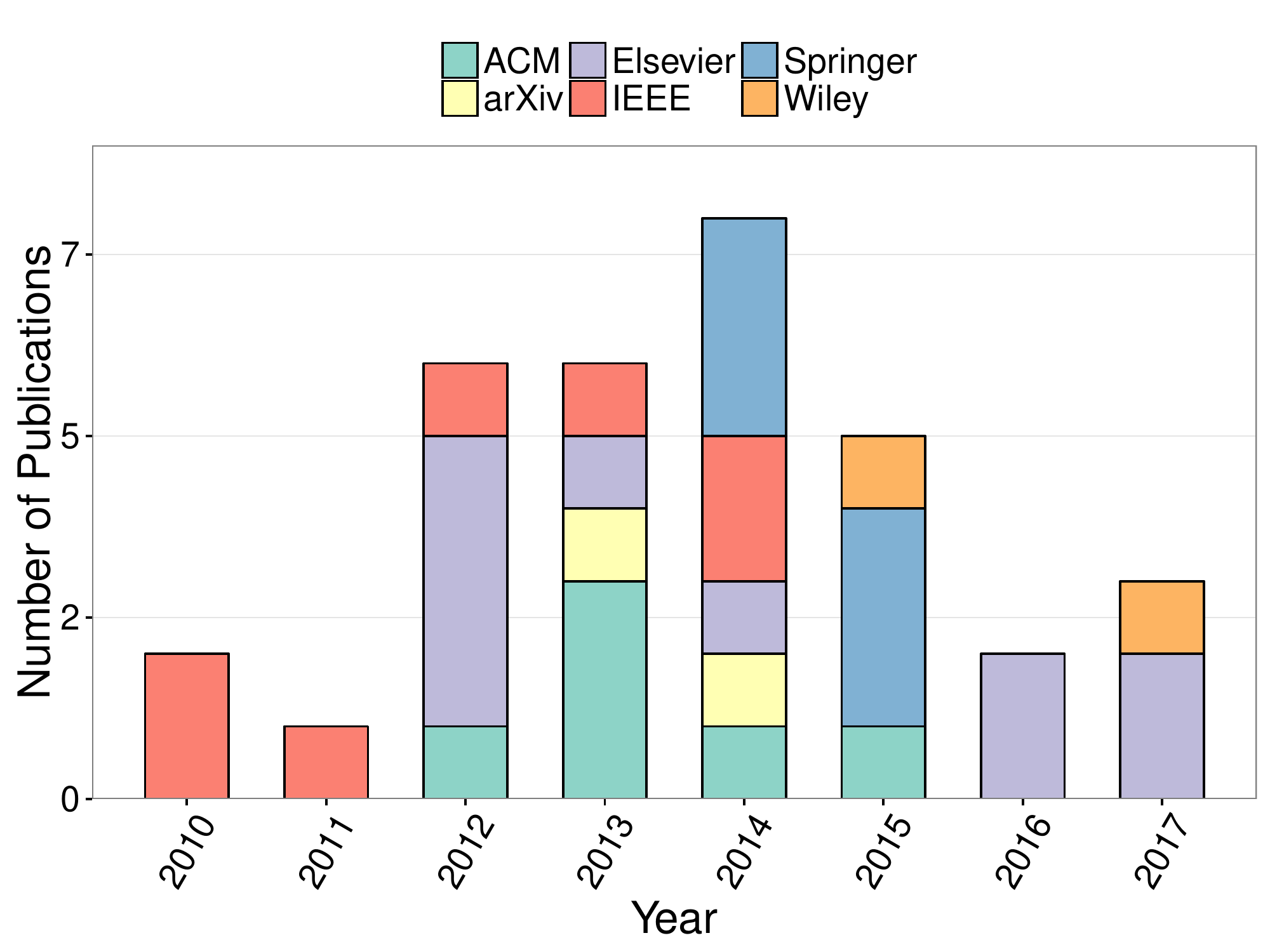}
\caption{Distribution of selected studies by publisher over publication year.}
\label{fig:results:publisher}
\end{figure}

The survey eventually identified a set of 33 principal papers on cloud brokerage solutions. As shown in \figurename~\ref{fig:results:pubtypes}, a significant proportion of these papers were published in journals ($\sim 48.5\%$), followed by a smaller proportion of publications ($\sim 33.3\%$) in conferences, approximately $12\%$ in workshops, and $6\%$ in other venues. 
When these results are plotted against publication year (\figurename~\ref{fig:results:pubyear}), we discern the typical pattern of early work being published in specialist conferences such as CCGrid, followed by later works in journals. 
The spread of original publishers (\figurename~\ref{fig:results:publisher}) is dominated by Elsevier and Springer; the former due to many papers being published in their FGCS journal whose call for papers was one of the earliest seekers of advancements in cloud brokerage, and the latter due to their pattern of publishing specialist conference proceedings. 

With respect to the reusability of the developed systems, the source code of 7 out of the 33 studies are made available to the public; 6 of them are published online and 1 is mentioned to be available upon request.

\section{Motivation for designing cloud brokers (RQ1)}
\label{sec:moivation}
This section looks at the challenges that motivated the cloud community to design cloud brokers and shows how selected papers addressed these challenges.
The majority of papers clearly state their motivation. From this, we found that cloud brokers are motivated by the need to address one or more of four key challenges:
\begin{enumerate*}[label=(\roman*)]
    \item Dimensionality, 
    \item Vendor lock-in, 
    \item Meeting requirements, and 
    \item Pathological.
\end{enumerate*}

\begin{figure}[htb]
\centering
\includegraphics[width=\textwidth,trim= 0 1cm 0 0]{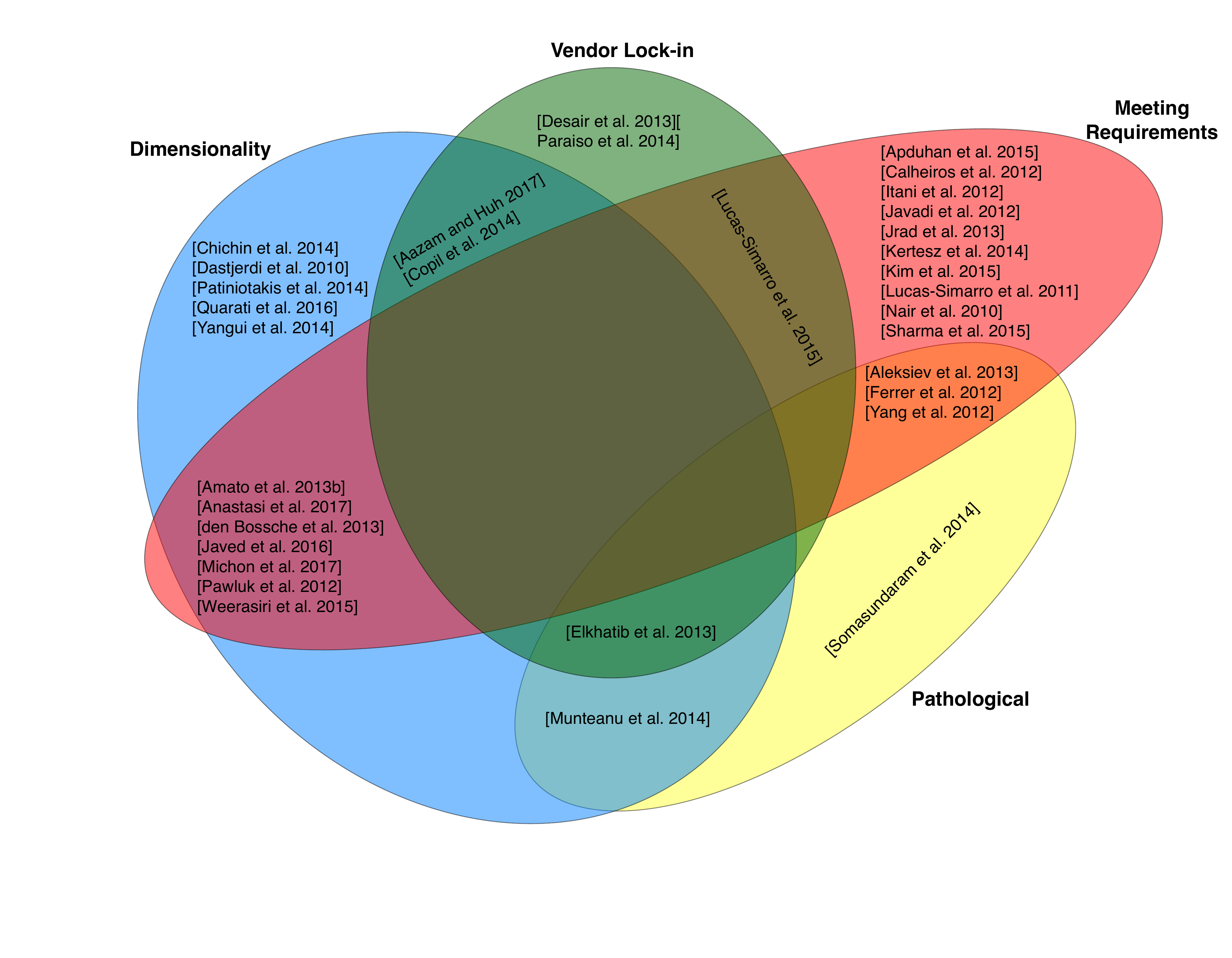}
\caption{The four main motivation categories and the papers in each.}
\label{fig:motivenn}
\end{figure}

\figurename~\ref{fig:motivenn} depicts where the surveyed papers lie across these different motivations. Most papers declare one or two motivations, with the need to tackle dimensionality and meet operational requirements being the most common. More than a third of the papers declare two or three motivations. None declare all four.

We now describe each category of motivation in detail.

\subsection{Dimensionality} 
\begin{figure}[htb]
\centering
\includegraphics[width=\textwidth]{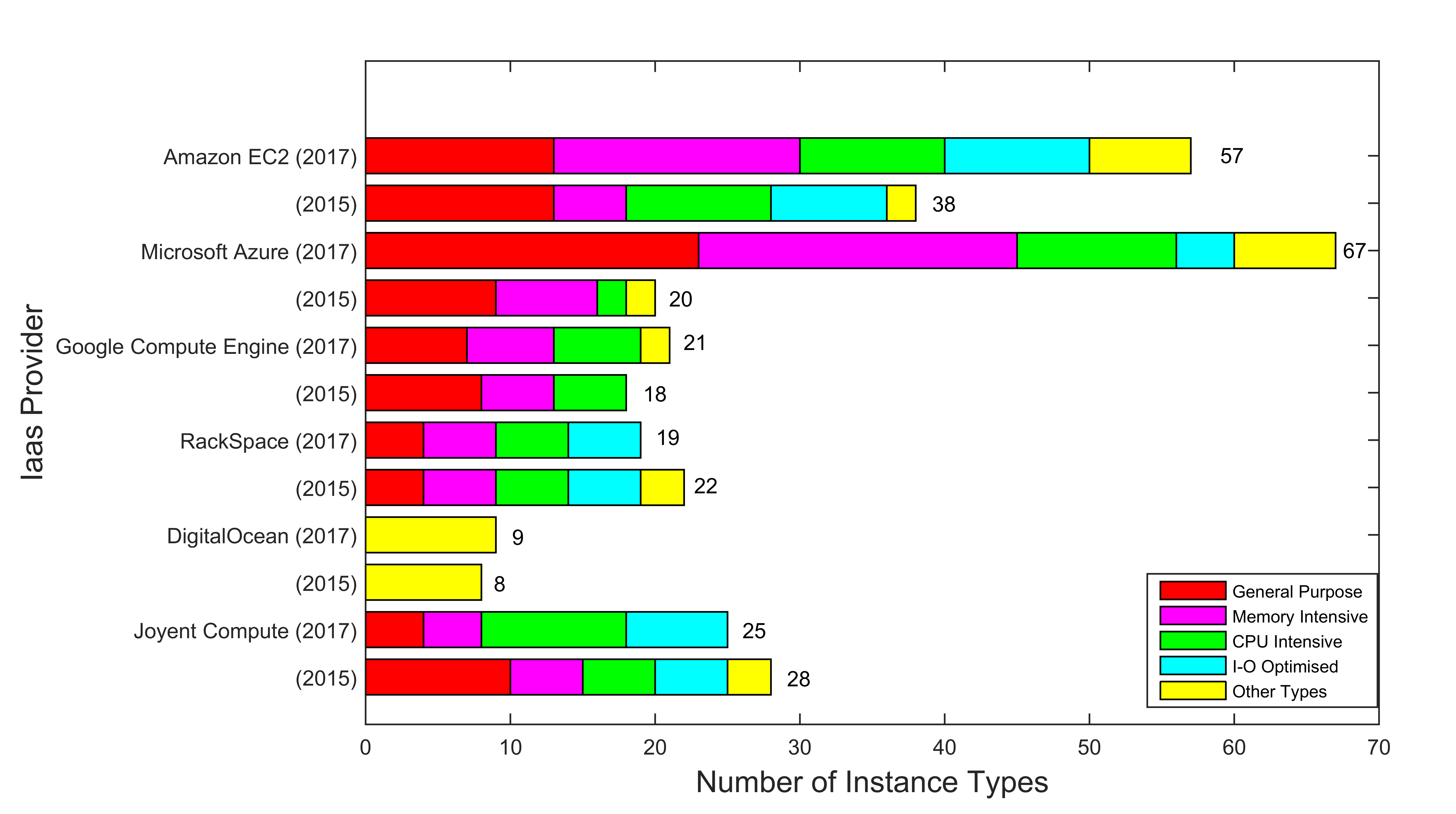}
\caption{The number of on-demand Linux cloud instance types from major CSPs in August 2015 and July 2017.}
\label{fig:noinstances}
\end{figure}

The vast number of cloud providers and their respective offerings makes the task of selecting a provider and service challenging. \figurename~\ref{fig:noinstances} shows the wide range of instance types offered in the IaaS market, and it illustrates how this wide range increased further between 2015 and 2017. In view of this, the current process of manually selecting the optimal service can overwhelm a human decision maker. This challenge motivates the development of systems to support decision making \cite{5493487,6928211,Patiniotakis2014PCS,VandenBossche2013,Pawluk2012stratos,Quarati2016403,Weerasiri2015,Yangui2014,Javed201652,Amato2013mosaic,Aazam2017,MICHON201711,ANASTASI20171}. 

\subsection{Vendor Lock-in} Vendor lock-in refers to the difficulty and expense that customers must sometimes incur to switch their service provider \cite{Leavitt2009,Hsu2014}. These costs and difficulties arise due to the technical differences between service providers, such as incompatibilities between CSP services and the financial cost of migrating large files. \citet{Desair2013} state that vendor lock-in limits the general exploitation of cloud computing due to the partial loss of control over data and applications. \citet{ParaisoMS14} are motivated by the need to develop CSP-independent applications, which is currently unfeasible due vendor lock-in. 
    
\subsection{Meeting Requirements} Many of the works we surveyed cite the inability of a single CSP to satisfy user requirements as a key motivator. For instance, \citet{Calheiros20121350} motivates the need for brokerage to enable applications to scale over multiple cloud data centres when the resources of one data centre are oversubscribed. \citet{Jrad2013} are motivated by the need to run workflow applications \footnote{Workflow applications are ``parallel applications that consist of a series of computational tasks logically connected by data- and control-flow dependencies'' \cite{Jrad2013}.} on multiple clouds. \citet{Kim2015} offers a similar motivation in the context of to mobile clouds. Additionally, observing the real performance of cloud services and assuring service level agreements (SLAs) motivates the need for brokers in \cite{Ferrer201266,Itani2012,Kertesz201454,ANASTASI20171,MICHON201711} to help customers select between different resources provisioning strategies, and decide on optimal VM placement \cite{Simarro2011,lucsimaro2015}. The provision of availability and reliability guarantees at low cost also motivate many works \cite{Javadi2012,Sharma2015spotcheck,Apduhan2015}. A recurring example is that of cloudbursting, \ie exploiting public cloud resources only when a situation arises where private cloud resources are insufficient or inadequate to satisfy application needs. This requires brokers to facilitate the addition of resources to scale beyond the limits of owned infrastructure by seamlessly adding and removing rented resources \cite{Nair2010towards,Aleksiev2013vmmad}. \cite{Copil2014controlling} argues that the existence of different APIs of the various providers motivates the need for a controller that provides complete view of the multi-cloud services so that users can have elastic control of their applications.
    
\subsection{Pathological Motivations} These motivations vary between achieving specific interoperability guarantees (\eg between different grid and cloud resources \cite{Aleksiev2013vmmad,Somasundaram2014,Munteanu2014interfacing}), enabling collaboration between users from different backgrounds \cite{elkhatib2013experiences}, hosting compute-intensive applications \cite{Yang2012}, and addressing complex legislative issues \cite{Ferrer201266}.

\section{Cloud broker functionality (RQ2)}
\label{sec:functionalities}
This section addresses RQ2 which looks at the functionalities that are performed by the cloud brokers. The following functionalities were extracted from the selected papers:

\begin{enumerate}
    \item \textbf{Decision Support.} Approaches implementing this functionality support the CSCs in making decisions concerning selection of resources. These approaches require that the CSC specifies functional and non-functional requirements to enable the system to recommend cloud providers and instances that best fit their needs. Users ultimately make the decision and allocate resources. Examples of systems implementing this functionality include \cite{Javed201652,6928211,Patiniotakis2014PCS,Jrad2013,Quarati2016403}. 
    
    \item \textbf{Resource Monitoring.} This concerns the ongoing collection of data from cloud services that could ultimately assist the broker in making effective decisions. Collected data generally includes metrics around performance and availability. This functionality can be implemented at the VM level as in \cite{Pawluk2012stratos}, the hardware level as in \cite{Yang2012}, or the application level as in \cite{Quarati2016403}.
    
    \item \textbf{Policy Enforcement.} User-defined policies provide constraints that guide selection and allocation decisions. Such constraints relate to the location of deployment, security, storage encryption and cost, among others. Brokers that implement policy enforcement implement and apply mechanisms to ensure that selected cloud services perform meet the constraints specified by the CSC. The broker may also use the policies to make adaptation decisions (\eg VM migration). 
    
    \item \textbf{SLA Negotiation.} SLA negotiation involves the broker capturing the user's requirements and enabling CSPs to tender cloud services that meet them. The result of the negotiation process is an SLA that defines commitments between the provider and the customer. SLA negotiation requires cooperation between CSPs and the broker to manage tenders and mechanisms to deal with SLA breaches. \cite{Nair2010towards,Ferrer201266,Aazam2017} are example of brokers that implement SLA negotiation. 
    
    \item \textbf{Application Deployment.} Application deployment is related to the place where that application will be made available. In the context of cloud brokerage, some brokers provide the functionality of deploying the application on the selected CSP service. Others do not support the actual deployment and just recommend a deployment environment.

    \item \textbf{Migration.} This addresses the challenges involved in changing the allocation of resources after resources are deployed. The need to migrate may be driven by changes in the CSC's requirements or by changes in the CSP's performance or availability. Six out of the selected studies support this functionality - all at the VM level.
    
    \item \textbf{API Abstraction.} Abstracting away technical differences between CSPs is fundamentally necessary to achieve seamless brokerage. This can be implemented in one of two ways. First, the broker may identify overlap between CSPs' APIs and present CSCs with the least common denominator (LCD). Alternatively, the broker may expose its own meta-API, calls to which are translated into native API calls. In both cases, the user must use the API exposed by the broker, that is likely to be substantially different to the CSPs' own APIs. The difference between these approaches lies in translation effort: the former prioritises ease of translation (ideally eliminating it completely) and thus results in a restrictive LCD API; the latter prioritises coverage over conformity.
    
    \item \textbf{VM Interoperability.} This functionality means that the broker allows the conversion of the VM (or the execution unit) between the different providers' formats. In \cite{Nair2010towards} interoperability is achieved through the standard API abstraction (OCCI) which is assumed to be adopted by all of the cloud providers. The work of \cite{5493487} (which applies ontology-based brokerage) introduces mediators that are defined as elements that handle interoperability between different ontologies or services.
\end{enumerate}

\section{Engineering approaches (RQ3)}
\label{sec:rq3}
This section addresses RQ3 which looks at the approaches proposed to design and construct the cloud brokers to highlight how they are realised. 

\subsection{Market-based}
Approaches in this category model proposed brokers as markets in which providers publish their services and consumers bid for  them. In this model, economic models form the matchmaking process between providers and consumers. The \textit{Smart Cloud Marketplace} (SCM) \cite{6928211} architecture consists of agents (\eg market-agent, buyer-agent and seller-agent) that perform  the role of a cloud service \textit{exchange}, making intelligent decisions on behalf of customers and providers. SCMs implement a variety of trading protocols such as fixed-price markets, negotiation, tendering, and auctions. Another example is the \textit{SpotCheck} broker \cite{Sharma2015spotcheck} that re-sells resources from IaaS providers in a unified market with varying guarantees and pricing models. This approach nests VMs within spot instances and migrates them during price spikes. In catering to spot instances, a service currently only offered by Amazon EC2, this broker uniquely relies on a single CSP. \citet{Javed201652} present a Cloud Market Maker (CMM) to increase providers' return on investment using a dynamic pricing strategy and to support CSCs in selecting CSPs. The CMM adapts the price of offered resources at operating-time using a supply-demand model that presents providers with the market equilibrium price, thereby encouraging providers to respond to demand.

\subsection{Frameworks and architectures}
The majority of work falls under this category where designed architectures that contain components, methods, and relationships are proposed. In some cases the authors provide implementations (either real or simulation-based using \eg CloudSim \cite{cloudsim2011}) for the framework while in others the frameworks remain at the conceptual levels. The proposed frameworks can be categorised into the following:

\subsubsection{Frameworks for executing applications on multi-clouds}
These frameworks focus on selecting multiple clouds to execute an application towards a better performance (\ie the application concurrently runs on multiple clouds). The application is divided into tasks so that each can be executed in a certain cloud. \citet{Jrad2013} implemented a framework for deploying workflow applications on inter-cloud environment, assuming that a standard API is adopted by the providers. The framework consists of a match-maker component that selects the data centres that satisfy the entire user predefined functional and non-functional requirements. It also consists of a scheduler component to distribute the workflow using the round robin scheduling policy. The brokerage system of \citet{Quarati2016403} also addresses the execution of workflow applications on grids and clouds. The system includes two main components. The first is the CB-Portlet which contains a number of portlets to enable the users to specify their workflow requirements. The second is the DCI-BRIDGE that enables the interoperability between the different computing resources (cloud or grid-based)  by creating the descriptions of the jobs that will be executed on the selected VMs; assuming the OCCI standard is adopted.
The framework of \citet{Copil2014controlling} is a simple one controller component for an application deployed across multiple clouds. The controller builds a dependency graph model to represent vendor-specific elasticity capabilities. It enables the user to specify requirements (regarding monitoring, elasticity constraints, strategy directives) at different levels (\eg service topology, code region). These requirements are then analysed by a runtime and an action plan is generated accordingly to fulfil the user's requirements.
\citet{Yang2012} proposed a framework to enable public clouds to cooperate to satisfy the infrastructure demands of CPU-intensive applications. The framework includes a coordinator that guarantees an SLA with hosts and provides the user with a token. This token enables the user to deploy to the host directly. The SLA is monitored and enforced by migrating some of the workload to another CSP. The \textit{soCloud} framework \citet{ParaisoMS14} extends the OASIS SCA\footnote{\url{http://www.oasis-opencsa.org/sca}} standard to build multi-cloud PaaS applications with added policy to ensure load balancing and high availability.The brokerage part is simple; it implements logic to ensure availability and minimum QoS. The application has to be written in a component-based fashion using the FraSCAti implementation of SCA. Negotiation among CSPs and pricing model variances are not considered.

\subsubsection{Frameworks for elastic application execution}
These frameworks focus on scaling up the cloud resources when the application's demands change. The VM-MAD framework \citet{Aleksiev2013vmmad} is specifically proposed for cloudbursting linux-based High-Performance Computing clusters. When the queue of a HPC cluster is overloaded, then cloud VMs are spun up to carry some of the load. Otherwise, cloud VMs are shut down when load recedes. 
\citet{Calheiros20121350} also proposes a cloud coordinator that manages the purchase of resources when changes in the application demands occur. The coordinator is assumed to be part of each CSP architecture and provides modules to enable CSPs to negotiate and exchange resources among themselves to satisfy elastic applications. 

\subsubsection{Frameworks for scheduling and resource allocation}
These frameworks focus on selecting the cloud resources based on application requirements and in some cases to increase the broker's revenue. 
\citet{Aazam2017} propose a framework to predict, reserve and allocate cloud resources. The framework is intended to be managed by a third-party for-profit body (\ie not to be user controlled). The authors propose algorithms to deploy the optimal services using a prediction of future demand (to maximise the profit of the third party). 
The framework of \citet{Kertesz201454} combines negotiation, brokering and deployment. The role of the broker is limited to managing the virtual resources. The broker functionality includes meta-brokerage and brokerage. The role of the meta-broker is to decide which broker is capable of satisfying the user requirements. On the other hand, the broker interacts with virtual and physical resources. 
\citet{Simarro2011,lucsimaro2015} divide the cloud broker into three components: the VM manager, the scheduler and the cloud manager. The role of the scheduler is to decide about the VMs placement among the available clouds. The cloud manager addresses the monitoring and management of the VMs life cycle. The authors focus on the scheduler component which is realised by algorithms that optimise the cost of required resources, based on the historical prices. 
In \citet{Javadi2012} the framework has three main components, namely, ``InterGrid Gateways (IGGs)'', the ``Virtual Infrastructure Engine (VIE)'' and ``Distributed Virtual Environment (DVE)'' manager. The IGG selects a suitable provider that for an incoming request. The VIE manages the private cloud resources (starts, pauses, resumes, and stops VMs) while the DVE allocates and manages resources on behalf of applications. 
The \textit{STRATOS} framework \cite{Pawluk2012stratos} provides automated decision making for resource acquisition between different vendors based on two steps: determining the number of resources required, and determining where to place the resources. 

\subsection{Toolkits}
In the context of cloud brokerage, toolkits can be defined as development tools that are used to design and develop broker functionality. 
The OPTIMIS toolkit \cite{Ferrer201266} consists of a set of components that allow (in theory) a variety of architectures of multiple clouds (a broker is one of the architectures). The toolkit allows the broker to act as a CSC of the CSPs and as a CSP for the CSCs. The  toolkit enables aggregating resources from CSPs and provisioning them to the CSCs. The work of \citet{Nair2010towards} 
focuses on addressing security concerns in OPTIMIS. The toolkit in \citet{elkhatib2013experiences} enables a shared space to facilitate data access and exchange between multiple communities on a common environmental issue such as flooding \cite{Wilkinson2015left} and chemical pollution \cite{Greene2015biogeochem}. The tool enables collaborators to access shared data that are stored in federated clouds. The tool includes a resource broker (RB) that selects a cloud service that is appropriate for the type of computation required. The contribution of \citet{ANASTASI20171} is a tool that assists users in selecting the most cost-efficient resources to execute scientific applications. The paper addresses the lack of knowledge about the actual cost of running an application is uncertain before runtime. This means that measuring the impact of alternative execution strategies (i.e. using alternative resources) can be expensive. The proposed tool, called Schlouder, is a simulator that predicts the cost of executions under various strategies; supporting the user to make decisions. The authors demonstrate the effectiveness of the simulator by showing that cost predictions are accurate. However, they admit that accuracy depends on users' ability to accurately estimate the duration of each task, which is difficult in practice.

\subsection{Middlewares}
A middleware is a system that provides bridging and interoperability between various systems. In the cross-cloud context, a middleware provides interoperability between different cloud vendors. Few approaches introduced their cloud brokers as middlewares. The CompatibleOne \cite{Yangui2014} middleware allows developers to combine different cloud services provided by different providers. It provides an abstraction called CORDS (CompatibleOne resource description system) to model and manage cloud services on CompatibleOne. CORDS abstracts, adds to and maps to OOP, OCCI\footnote{\url{http://occi-wg.org/}}. The goal is to enable developers to specify a service in enough detail to enable it to be created at any CSP. \citet{Munteanu2014interfacing} also presents a middleware for interfacing with multiple IaaS and PaaS providers. The paper proposes a basic technique of SLA negotiation between more than one candidate provider. This is done using descriptors that are semantically matched to CSP capabilities. The presented descriptor-driven framework is VM-specific, relying on a brief SLA comparison phase followed by execution of predefined scripts. This is limited by the expressiveness of the SLAs provided by different vendors and the semantic differences therein. It also seems to exclude brokering other types of cloud provisions such as storage, appliances. PaaSHopper \cite{Desair2013} is a policy-driven middleware that attempts to mitigate the problem of losing control over application and data when adopting cloud resources. It includes two layers, namely, the abstraction layer and the policy-driven distributed execution layer. The former deals with heterogeneity and interoperability with a uniform API for interaction to middleware. The latter allows users to specify constraints to select the location of application execution (in the private or the public cloud) in addition to other criteria related to data security \eg connection security and storage encryption.

\subsection{Semantics}
These approaches deal with the heterogeneity of cloud resources' configuration and the cloud services' APIs. The general case aims at unifying the CSPs resource configuration annotations and common programming APIs. They propose ontologies to achieve this goal and assume that CSPs will adopt such common technologies. \citet{Weerasiri2015} propose a common resource configuration language, called ``Resource Configuration Service'' (RCS) which enables abstracting the heterogeneous CSPs services interfaces through a unified interface. The authors also propose a ``Cloud Resource Configuration Description''(CRCD), which enables provider-independent resource requirement descriptions. Such abstractions enable CSCs to model and execute cloud applications by focusing on specifying the applications' requirements. \citet{5493487} propose an ontology to model the virtual resources based on the Web Service Modeling Ontology (WSMO)\footnote{\url{http://www.wsmo.org}}. The ontology provides the domain concepts and vocabulary for representing resources advertisements and requestor requirements. 
\citet{Apduhan2015} propose a broker-server that resides on the private cloud resources within a hybrid cloud. The proposed broker-server is based on adopting a cloud ontology in order to search, rank and select cloud services. \citet{Somasundaram2014} propose a domain specific ontology that provides a mechanism to represent grid and cloud resources information. The ontology is used in a framework for resource discovery, SLA negotiation, and resource scheduling. 

\subsection{Algorithms}
Some works focus on proposing algorithms to realise (in some cases partially) the brokerage system as a main contribution. \citet{Amato2013mosaic} propose a distributed algorithm where the broker is decomposed into a set of distributed brokers (each resides at a CSP) that cooperate to satisfy the users requests. The global broker sends out a call for proposals to satisfy the user's requirements, that include quantitative specifications as defined in OCCI and qualitative factors such as location. Each cloud runs an agent that offers at least one proposal to the broker to satisfy the user's requirements. The global broker then collects the proposals and selects the most appropriate one according to a brokering policy. \citet{Itani2012} propose an algorithm for brokerage in federated clouds inspired  by the BGP routing protocol. Public clouds are advertised as if they were an autonomous systems in BGP. The broker routes procurement requests to one of the clouds according to its policy. The CSP either satisfies the request entirely or satisfies it transitively. For example, if the user requests a database server, the first CSP may request a storage server from another CSP as a prerequisite. \citet{Kim2015} propose a two-phases algorithm for workflow execution in mobile cloud, namely, the ``greedy based workflow co-scheduling (GBWC)'' phase and the ``resource profiling based placement (RPBP)'' phase. This two-phase algorithm is then used by the ``Mobile Cloud Broker (MCB)'' in the proposed system for cost adaptive VM management. The GBWC reduces the cost of utilising resources by simultaneous scheduling of tasks on the same VMs. The RPBP algorithm selects cloud physical resources to host a VM  using simple profiling algorithms. The profiling is based on either the CPU capacity or the historical network delay. The profiles are used to rank the resources then the highest rank resources are selected subject to the budget constraints. 
\citet{VandenBossche2013} propose an algorithm to schedule deadline-constrained workloads on a hybrid cloud. A hybrid cloud scheduler decides (based on the budget and deadline) whether a task can be scheduled on the private cloud or if it should be transferred to the public cloud. The tasks are stored in a queue based on their deadlines. A queue scanning algorithm is used to detect tasks that cannot be satisfied within the deadline (based on the elapsed and remaining time of the tasks in the queue). These tasks will then be executed on the public cloud. In \citet{Patiniotakis2014PCS}, a  cloud service recommender (called PuLSaR) is proposed as a multi-criteria decision making approach to compare and recommend cloud services to customers. The selection is based on the customer's preferences towards the services metrics. The recommender system uses the notion of imprecise metrics along with precise metrics to capture the fuzzy or linguistic values provided as requirements by the customer. Then a fuzzy analytical hierarchical process is used for comparison and ranking. In \citet{MICHON201711}, the QBrokerage algorithm is proposed based on a genetic algorithm to select cloud providers. First, the cloud application is modeled as a directed graph in which the vertices represent virtual machines and edges represent the communication paths between virtual machines. Second, the graph is passed to a mapping component which searches for services that satisfy the requirements of the VM.

\section{Characterisation and Discussion of the State of the Art}
\label{sec:chr}
We now reflect on the selected studies as a representation of the state of the art in cloud brokerage. We first compile and summarise the above results through a characterisation framework then we present general observations and discuss a number of limitations we identified in the literature. 

\subsection{Characterisation Framework}
We now give an overall characterisation of the selected studies in Table~\ref{tab:res:summ}. To achieve this, we identified a framework of 15 main characteristics that differentiate the surveyed works. These are as follows: 

\newcounter{magicrownumbers}
\newcommand\rownumber{\stepcounter{magicrownumbers}\arabic{magicrownumbers}}
\begin{table}[htbp]
\caption{Summary of accepted papers.\label{tab:res:summ}}{
	\def\arraystretch{1}
	\setlength{\tabcolsep}{1.8pt}
	\rowcolors{2}{blue!17}{white}
	\newcommand{\citex}[1]{\scriptsize{\cite{#1}}}
	\begin{tabular}{rlccccccccccccccccc}
 		& Reference & \vrt{Motivation} & \vrt{Provisioning Level} & \vrt{Comparison Framework} & \vrt{API Abstraction} & \vrt{Bidding Style} & \vrt{Monitoring Level} & \vrt{Policy Enforcement} & \vrt{SLA Negotiation} & \vrt{Network Costed} & \vrt{Application Deployment} & \vrt{Migration} & \vrt{VM Interoprability} & \vrt{Lifecycle Management} & \vrt{Evaluation} & \vrt{Source Available} \\
		\hline
 		\rownumber & \citex{5493487} & D & I & WSMO & \xmark & Q & A & \xmark & \xmark & \xmark & \cmark & \xmark & \cmark & \xmark & P & \xmark \\
 		\rownumber & \citex{Nair2010towards} & M & I & \xmark & Own & Q & V & \cmark & \cmark & \xmark & \cmark & \cmark & \cmark & \cmark & P & \href{http://optimistoolkit.com/}{\cmark} \\
 		\rownumber & \citex{Simarro2011} & M & I & \xmark & deltacloud & C & V & \xmark & \xmark & \xmark & \cmark & \xmark & \xmark & \xmark & S & \xmark \\
 		\rownumber & \citex{Ferrer201266} & MP & I & \xmark & Own & Q & V & \cmark & \cmark & \xmark & \cmark & \cmark & \cmark & \cmark & P & \href{http://optimistoolkit.com/}{\cmark} \\
 		\rownumber & \citex{Itani2012} & M & I & \xmark & \xmark & C & \xmark & \cmark & \xmark & \xmark & \xmark & \xmark & \xmark & \xmark & S & \xmark \\
 		\rownumber & \citex{Pawluk2012stratos} & DM & P & SMI & deltacloud & C & V & \cmark & \xmark & \xmark & \cmark & \xmark & \xmark & \cmark & R & \href{http://stratos.apache.org/}{\cmark} \\
 		\rownumber & \citex{Javadi2012} & M & I & \xmark & \xmark & D & V & \xmark & \xmark & \xmark & \cmark & \xmark & \xmark & \xmark & S & \xmark \\
 		\rownumber & \citex{Yang2012} & MP & H & \xmark & \xmark & Q & H+A & \cmark & \cmark & \xmark & \cmark & \cmark & \xmark & \cmark & P & \xmark \\
 		\rownumber & \citex{Calheiros20121350} & M & I & \xmark & \xmark & C & V & \xmark & \cmark & \xmark & \cmark & \xmark & \xmark & \xmark & R+T & \xmark \\		
 		\rownumber & \citex{elkhatib2013experiences} & VMP & I & Own & jclouds & Q & V & \cmark & \xmark & \xmark & \cmark & \xmark & \xmark & \cmark & P & \xmark \\
 		\rownumber & \citex{Jrad2013} & M & I & OCCI & \xmark & Q & V & \xmark & \xmark & \xmark & \xmark & \xmark & \xmark & \xmark & S & \xmark \\
 		\rownumber & \citex{Desair2013} & V & P & \xmark & Own & C & A & \cmark & \xmark & \xmark & \cmark & \xmark & \xmark & \xmark & P & \xmark \\
 		\rownumber & \citex{Amato2013mosaic} & DM & S+P+I & OCCI & \xmark & MBCD & \xmark & \xmark & \cmark & \cmark & \xmark & \xmark & \xmark & \xmark & S & \xmark \\
		\rownumber & \citex{Aleksiev2013vmmad} & MP & I & \xmark & \xmark & Q & V & \cmark & \xmark & \xmark & \xmark & \xmark & \xmark & \cmark & T & \href{https://github.com/uzh/vm-mad}{\cmark} \\
 		\rownumber & \citex{VandenBossche2013} & DM & I & \xmark & \xmark & D & V & \xmark & \xmark & \xmark & \cmark & \xmark & \xmark & \xmark & T & R \\
 		\rownumber & \citex{Kertesz201454} & M & I & Own & \xmark & Q & A & \xmark & \cmark & \xmark & \cmark & \xmark & \xmark & \xmark & S & \xmark \\
 		\rownumber & \citex{ParaisoMS14} & V & S+P & OASIS SCA & FraSCAti & Q & A & \cmark & \xmark & \xmark & \cmark & \xmark & \xmark & \cmark & R+T & \xmark \\
 		\rownumber & \citex{Munteanu2014interfacing} & DP & P+I & Own & mOSAIC & M & \xmark & \cmark & \cmark & \cmark & \cmark & \cmark & \xmark & \xmark & P & \href{https://bitbucket.org/mosaic/}{\cmark} \\
 		\rownumber & \citex{Copil2014controlling} & DM & P+I & Own & \xmark & C & A & \xmark & \xmark & \xmark & \cmark & \xmark & \xmark & \xmark & P & \xmark \\
 		\rownumber & \citex{Somasundaram2014} & P & I & Own & \xmark & D & V & \xmark & \cmark & \xmark & \cmark & \xmark & \xmark & \xmark & T+S & \xmark \\
 		\rownumber & \citex{Patiniotakis2014PCS} & D & U & SMI+USDL & \xmark & C & \xmark & \cmark & \xmark & \xmark & \xmark & \xmark & \xmark & \xmark & P & \xmark \\
 		\rownumber & \citex{6928211} & D & U & \xmark & \xmark & M & \xmark & \cmark & \cmark & \xmark & \xmark & \xmark & \xmark & \xmark & S & \xmark \\
 		\rownumber & \citex{Yangui2014} & D & P+I & OCCI & Own & C & V & \cmark & \cmark & \xmark & \cmark & \cmark & \cmark & \cmark & P & \xmark \\
        \rownumber & \citex{lucsimaro2015} & VM & I & \xmark & deltacloud & C & V & \xmark & \xmark & \xmark & \cmark & \xmark & \xmark & \xmark & S & \xmark \\
 		\rownumber & \citex{Kim2015} & M & I & \xmark & \xmark & BQ & V & \xmark & \xmark & \cmark & \cmark & \xmark & \xmark & \xmark & T & \xmark \\
 		\rownumber & \citex{Weerasiri2015} & DM & S+P & \xmark & Own & C & \xmark & \xmark & \xmark & \xmark & \cmark & \xmark & \xmark & \xmark & P & \href{https://github.com/ddweerasiri/Federated-Cloud-Resources-Deployment-Engine}{\cmark} \\
 		\rownumber & \citex{Apduhan2015} & M & I & Own & \xmark & B & V & \xmark & \xmark & \cmark & \xmark & \xmark & \xmark & \xmark & S & \xmark \\
 		\rownumber & \citex{Sharma2015spotcheck} & M & I & \xmark & Own & M & V & \cmark & \xmark & \cmark & \cmark & \cmark & \xmark & \cmark & P+S & \xmark \\
 		\rownumber & \citex{Javed201652} & DM & I & \xmark & \xmark & M & \xmark & \xmark & \cmark & \xmark & \xmark & \xmark & \xmark & \xmark & S & \xmark \\
 		\rownumber & \citex{Quarati2016403} & D & I & OCCI & DCI Bridge & B & A & \cmark & \xmark & \xmark & \cmark & \xmark & \xmark & \cmark & S & \xmark \\
 		\rownumber & \citex{Aazam2017} & V & U & \xmark & Own & B & \xmark & \xmark & \cmark & \xmark & \cmark & \xmark & \xmark & \cmark & S & \xmark \\
 		\rownumber & \citex{MICHON201711} & DM & I & \xmark & \xmark & Q & V & \cmark & \xmark & \xmark & \cmark & \xmark & \xmark & \xmark & R+T & \href{http://schlouder.gforge.inria.fr/}{\cmark} \\
 		\rownumber & \citex{ANASTASI20171} & DM & I & \xmark & \xmark & Q & \xmark & \xmark & \xmark & \xmark & \cmark & \xmark & \xmark & \xmark & S & \xmark \\
 		\hline
	\end{tabular}
}
\end{table}

\begin{enumerate}
\item \emph{Motivation} -- the motivation for proposing a cloud broker: dimensionality (D), vendor lock-in (V), meeting requirements (M), or pathological motivations (P).
\item \emph{Provisioning Level} -- the type of resources that the system brokers: HaaS (H), IaaS (I), PaaS (P), SaaS (S), or unspecified (U). 

\item \emph{Comparison Framework} -- if a standard index or ontology is used to differentiate cloud vendors (\eg OCCI), if the authors developed their own ontology (Own), or none is adopted (\xmark).

\item \emph{API Abstraction} -- whether the broker exposes a unified API with which developers can control resources at CSPs,  (\eg libcloud), their own (Own), or not (\xmark).

\item \emph{Bidding Style} -- refers to the strategy used by the broker to select resources, and it could be:
\begin{itemize}
	\item Resource-oriented (C) where the user indicates a minimum specification of computational resources;
	\item Deadline-driven (D) where the user provides a target for executing a set of jobs within, possibly in addition to minimum computational resource requirements; 
	\item QoS-oriented (Q) where the user is more interested in maintaining certain levels of QoS rather than total execution time; 
	\item Budget-based (B) where monetary or energy budgets are declared or identified, which the scheduler needs to remain within (\ie hard budget) or tries to remain within (\ie soft budget), possibly in addition to minimum computational resource requirements; 
	\item Resale (R) where the broker buys spare resources from vendors and resells to the users; 
	\item Marketplace (M) where both providers and users place their bids and the broker matches the available buying and selling prices.
\end{itemize}

\item \emph{Monitoring Level} -- the monitoring granularity: hardware level (H), VM level (V), appliance or service level (A), or none (\xmark).

\item \emph{Policy Enforcement} -- if the broker ensures the execution of user-defined resource procurement and management policies (\cmark) or not (\xmark).

\item \emph{SLA Negotiation} -- if the broker offers any mechanisms to negotiate SLA specifics with the vendor on behalf of the user (\cmark) or not (\xmark).

\item \emph{Network Costed} -- whether the cost of deploying to a new cloud provider include the network latency and bandwidth utilisation costs incurred (\cmark) or not (\xmark).

\item \emph{Application Deployment} -- whether the broker handles deployment to different cloud providers (\cmark), or if it merely acts a consultation service that offers recommendations (\xmark).

\item \emph{Migration} -- whether the broker provides support for migrating between different cloud providers (\cmark), or not (\xmark).

\item \emph{VM Interoperability} -- if the broker offers tools to convert VMs or other execution units between different vendors  (\cmark), or not (\xmark).

\item \emph{Lifecycle Management} -- if the broker directly manages the creation, maintenance and graceful destruction of VMs or other execution units (\cmark), or not (\xmark).

\item \emph{Evaluation} -- how the proposed solution is tested: on a number of real providers (R), in a testbed / private cloud (T), using simulation / replay of metrics (S), or a prototype / proof of concept (P).

\item \emph{Source Available} -- if the source code is released under an open source license (\cmark with a hyperlink), available upon request (R), or non-specified (\xmark).

\end{enumerate}

\subsection{General Observations}
In characterising the selected studies, we observe that:

\subsubsection{Interoperability is a concern only recently addressed}
Early brokerage attempts focused mainly on the relatively simple challenge of cloudbursting (\ie saturate private cloud resources before adding public cloud resources if/when necessary) \eg \cite{Javadi2012,Aleksiev2013vmmad,VandenBossche2013}. Other early attempts employed rudimentary selection mechanisms, using round robin and similarly na\"ive techniques, \eg \cite{Itani2012}. Moreover, many were scheduling oriented and deadline driven and did not implement policy enforcement or SLA negotiation.

\subsubsection{Basic bidding}
To further elaborate on the previous point, we look into the distribution of bidding styles. Most works (12 $\approx$36\%) follow a QoS-oriented model. 
This is closely followed by na\"ive resource-oriented methods (11 $\approx$33\%) that offer a minimum resource provisioning level. 
More sophisticated models such as Marketplace are only observed in more recent works. The Resale bidding style is surprisingly missing from the selected works. We have come across this approach in a number of works that did not make it past the screening phase mainly because they are designed for homogeneous cloud federations.

\subsubsection{IaaS is the most common} Most work focuses on brokerage at the IaaS level, with a few works also including PaaS and/or SaaS, and only one solution to brokerage for HaaS. This is unsurprising for a few reasons. 
First, IaaS offers the most flexibility in terms of software and hardware resources. Typically, the user selects from a range of operating system templates and hardware configurations and is responsible for configuring the system themselves. Because operating system templates and hardware specifications are relatively homogeneous among CSPs, offers can be compared simply and accurately. For example, it is trivial to identify offers with Ubuntu Server 16.04, 2 cores and 4GB RAM.

Second, the level of control afforded to IaaS users empowers them to migrate between service providers more easily.

Third, brokering at the IaaS level offers the most room for economic benefit. Resources provided at higher levels come at a premium and as such the margins for reducing cost across providers becomes thinner at PaaS and SaaS levels. Monitoring is also mostly done at the IaaS level. This again relates to comparability between like and like across providers. As such, the most common level of granularity is the VM.

\subsubsection{Standards and common libraries are not commonly used} Most papers do not use an ontology to compare cloud service offerings. Those that do tend not to follow standards. The most prevalent and mature ontology, OCCI, is only employed by four studies, while six create a bespoke comparison framework. 
The lack of enthusiasm to implement standards is not restricted to service comparison frameworks; API abstraction frameworks including common open source libraries, \eg libcloud and jclouds, are not at all widely adopted. Despite their pitfalls (\cf \cite{Elkhatib2016crosscloudmap}), these libraries are fairly mature by the time of the majority of these works are published. As such, this reinventing of the wheel is surprising as some authors choose to develop very similar API abstraction libraries that have no discernible benefits over open source community efforts.

\subsubsection{Different interpretations of delegation} The characterisation framework provides a comprehensive overview of the challenges deemed most pressing in cloud brokerage. It is clear that deployment is a central functionality for most works: 25 papers ($\approx$76\%) provide a solution to deploy VMs or appliances. This seems to indicate that delegation is a key priority. However, many of these works fail to provide delegation beyond deployment. For instance, only 11 works ($\approx$33\%) provide post-deployment life cycle management, and only 6 ($\approx$18\%) tackle migration. 

\subsubsection{The legacy of the grid} There is a significant number of works that either explicitly or implicitly emerged from previous projects and research groups focussing on grids and HPC computing infrastructures (\eg \cite{Javadi2012,Somasundaram2014}). It is, however, unfortunate that the majority of these works are mere reapplication of old technology with little consideration of the different nature of the cloud resources, such as access policies and management strategies. 

\subsubsection{Network blindness} Network-related aspects are often ignored in the literature: only 5 papers ($\approx$15\%) consider network costs. 
It is worth noting that only 2 of the works that propose migration support incorporate network costs into their solutions.

\subsubsection{Limited real-world tests} The majority of works rely on simulations or prototypes for verifying their solutions. This could be attributed to the costs associated with evaluating on public cloud resources and to ease of use of widely used simulators such as CloudSim. 

\subsection{Critical Reflections on the State of the Art}

By reflecting on the state of the art in both its details (Sections~\ref{sec:moivation}--\ref{sec:rq3}) and using the characterisation framework (Table~\ref{tab:res:summ}), we identify a number of limitations that we now discuss.

\begin{itemize}
    \item \textbf{Interoperability support.} Little has been done to tackle interoperability challenges. This is evidenced by the sparse adoption of mapping and translation libraries to abstract away differences between service providers' APIs. In addition, many frameworks assume that CSPs will adopt a common API so that they can collaborate to satisfy users' requirements. In reality, the cloud market is has demonstrated its resistance to allowing customers to freely move between competitors. 
    
    \item \textbf{Comparability approaches.} The vast majority of literature assumes that data published by CSPs is comparable. However, it is obvious that the services offered are fundamentally heterogeneous; each CSP has its own description of the offerings. Similar to the above, some approaches assume that the CSPs will adopt standard offering descriptions (\eg OCCI) which is also unrealistic. 
    
    \item \textbf{Adaptive deployment.} Very few frameworks handle deployment and fewer still handle migration. This is, perhaps, unsurprising as such operations are non-trivial. However, this reduces the majority of the contributions to consultancy services rather than fully developed brokers. Furthermore, the approaches that support migration do it at the `heavy' VM level. No brokerage solution currently takes advantage of the migration capabilities offered by container technologies \cite{hadley2015multibox}.
    
    \item \textbf{Intelligence.} The selection of CSPs and instances is based only on matching services to requirements. In other words, the selection is based on what the CSPs promise to provide not on the actual provisioning of resources. Recent analysis of some cloud instances show surprising results of inconsistent performance of the promised offerings. For example, as we have reported in \cite{Samreen2016Daleel}, the performance of Amazon \texttt{c4.xlarge} instance is quite the same as the performance of \texttt{c4.large} although the former is twice both in specification and cost of the latter. This realistic performance and cost issues cannot be observed by basing the selection only on the offerings' specifications, which can result in inability to meet the users' requirements or/and `unfair' costs.
    
    \item \textbf{Customer assistance.}
    The vast majority of the research assumes that CSCs are acutely aware of their applications' technical requirements \textit{before} deployment. In fact, it is very difficult for users to ascertain the necessary hardware specifications in advance \cite{Lango2013}. This is particularly true when demand for the application is expected to be volatile, as is often the case when IaaS is chosen above bare metal.

    \item \textbf{Realistic evaluation.} Evaluation is carried out mostly using simulations or using testbeds made up of private cloud resources. Several papers suggested that this is due to the costs of using public cloud services. Indeed it does cost money to carry out benchmarks, develop use cases, and evaluate prototypes. However, we know from experience of our own experiments \cite{Samreen2016Daleel} that such costs in total are well below the price of purchasing comparable hardware that could be used as private cloud resources. Moreover, several CSPs (such as Amazon and Microsoft) offer programmes to fund researchers with credits to use their cloud services for research purposes. This observation, coupled with the fact that API interoperability and open source releases are ignored by many works, leads us to suspect that most works are merely proof of concept ideas that are of little practical use to other researchers and practitioners. In other words, the literature still lacks pragmatic brokers that are evaluated on real cloud infrastructures, and that are usable by others.
\end{itemize}

\section{Future Directions}
\label{sec:future}

Based on our investigation and reflection, we identify a number of future avenues in the field of cloud brokerage.

	\subsection{Customer Assistance} As already mentioned in the previous section, 
	little has been done to assist customers in specifying the necessary low level technical requirements that best suits their high level requirements. In this context, further work is needed to translate high level objectives into cloud service specifications. This can be achieved by, for example, developing domain specific languages (DSLs) that capture customers service level objectives (SLOs) and translate them into measurable operational goals. Another approach is to learn application requirements at runtime based on the actual performance of applications relative to the cloud services in which they reside. Learnt models can then be used to inform the low level requirements of similar applications. This also requires developing methodologies and measures to define and quantify the similarity of the cloud applications. Furthermore, a trace-and-replay approach is another option to assist in specifying the applications requirements \cite{CloudProphet2011}. However, reducing the costs of adopting such approach needs to be taken into consideration.

	\subsection{Adaptive and Fluid Deployment} Containers are lightweight execution units that provide an efficient alternative to `classical' VMs in terms of resource utilisation. The layer of abstraction offered by containers makes CSC-centric migration a promising approach to adapt applications deployment. However, this is challenged by the need for methodologies to benchmark the containers so that the heterogeneous resources can be compared and ranked \cite{Varghese2016}. Other challenges relate to the management of the container lifecycle in the cross-cloud context and quantifying the cost of migration, among others. 
	
	\subsection{Intelligent Decision Making} Currently, decision making systems rely heavily on simplistic views of application requirements and service provider performance.  Machine-learning techniques can help to quantify the extent to which providers fulfil their offerings and the extent to which the applications are satisfied in practice. Also, machine learning models can be deployed to adapt deployments proactively, i.e. before the adaptation becomes necessary. A major challenge here is to accurately predict when requirements will change and the feasibility adapting the deployment at that time. Furthermore, feedback on the performance of the adopted learning approaches is necessary as learning in different ways can lead to different results.

\section{Limitations and Threats to Validity}
\label{sec:discussion_limitations}
We followed the systemic approach of \cite{kitchenham2004procedures} to avoid selection bias rather than relying only on researchers' knowledge and background. As researchers from different communities and backgrounds tend to use different terminologies for the same topics, we performed the search using as many terms as we know that are related to cloud brokerage, and results from all publication databases have been inspected accordingly. 
Reflecting on this methodology, we identified the following limitations the study is potentially subject to. We describe our mitigation strategy associated with each.

\subsection{Sampling bias}
The initial search phase was based on publication meta-data (\ie abstract, title, and keywords). It is possible that studies that have addressed brokerage in cloud computing have been overlooked by this method if none of their meta-data mention our search terms. However, this is unlikely as our search terms were rather broad, as evidenced by the huge search return in Phase 1 (Table~\ref{tab:meth:phases}). Furthermore, as meta-data are provided by the authors of the selected works, we believe it is reasonable to rely on the quality of the classification and indexing of the papers by the publishing databases. 

\subsection{Reporting bias and construct validity}
Our reporting of the review was based on a set of predefined research questions and guided by a deterministic list of inclusion and exclusion criteria. The selected studies were reported in light of the research questions, using thorough understanding of the different works to answer the research questions and in turn untangle the overlap between the studies. Works were also discussed between the researchers carrying out the survey, reducing chances of misrepresentation and, subsequently, further reducing threats to construct validity.

\subsection{Data extraction and internal validity}
The labour-intensive nature of this research could potentially contribute to inaccuracy in data extraction. We designed our methodology (described in section \ref{subsubsec:screening}) to minimise the possibility of human error in the data extraction process. Our core methodology was complemented by extensive and profound discussions among this paper's authors to ensure coherent and correct extraction of data.

\subsection{Reproducibility}
The study could be easily replicated using the search and selection strategies we outlined. However, full reproducibility of our work hinges on automated means of data collection. We found automated data collection to be more difficult to obtain than we had expected. First, most online libraries do not expose public web APIs and some even forbid web scraping. For those that do offer APIs, the search capabilities are rather limited and divergent\footnote{We found arXiv's API to be the most advanced in terms of functionality and reliability of results as search terms change, with those of PLOS, ScienceDirect and Springer close behind.}. 
Some APIs are restricted to paying customers and partners only; \eg IEEE and Wiley. 
ACM's API was extremely rudimentary and limiting, requiring significant post-processing of the retrieved results. Consequently, investment into automated development is a significant effort.

\subsection{External validity}
Our survey was concerned with peer-reviewed academic works only, and is only generalisable for such type of brokerage works only. 
We did not attempt to capture industrial efforts that are not published through academic research channels, although we are aware of a number of such works. This decision was made in order to avoid systematic errors and maintain the quality at which the survey is conducted.

\section{Conclusion}
\label{sec:conc}

This study set out to build an understanding of different solutions in the burgeoning field of cloud brokerage. For this, we followed a systematic literature survey in order to ensure thorough coverage of such solutions. We aimed to accurately represent the state-of-the-art in cloud brokerage and identify key accomplishments and challenges. We considered seven main publication databases to perform the search. We considered 10,847 papers from which we found 33 papers exhibiting acceptable relevance to the survey topic. 

The main findings show that brokerage in the cloud emerged as a cross-cloud model that is motivated by the heterogeneity and dimensionality of the current cloud services in addition to the limitation of the single-cloud paradigm to satisfy the customers requirements. The proposed cloud brokers are intended to perform a number of tasks related to supporting customers' decisions making, application deployment, SLA negotiations, and resources monitoring, among others. These brokers have been engineered using diverse approaches including middlewares, toolkits, frameworks, and others. Section \ref{sec:rq3} has provided some details about the methods and
techniques used to clarify the ways in which brokers are realised. We also identified the limitations of the current cloud brokers. In general, we observed that most proposals implement a subset of the functionality typically associated with a broker. We see this normal as the field is still in its formative stage. The field needs further work to assist CSCs in specifying their applications' requirements, adaptation and intelligent decision making related to the selection of the cloud providers and services. Our future work will focus on addressing the challenges outlined in this survey.

\section*{Acknowledgement}
This work is partly supported by the Adaptive Brokerage for the Cloud (ABC) project, UK EPSRC grant EP/R010889/1.

\bibliographystyle{plainnat}
\bibliography{bibs/survey,bibs/extra,bibs/stage5}

\end{document}